\documentclass[aps,prb,twocolumn,superscriptaddress,showpacs]{revtex4-1}

\usepackage[pdftex]{graphicx}
\usepackage{amsmath}
\usepackage{color}
\usepackage[normalem]{ulem}

\begin{document}

\title{Generation of Rabi frequency radiation using exciton-polaritons}
\author{F\'abio Barachati}
\email[]{fabio-souza.barachati@polymtl.ca}
\affiliation{Department of Engineering Physics, \'Ecole Polytechnique de Montr\'eal, Montr\'eal H3C 3A7, QC, Canada}
\author{Simone De Liberato}
\affiliation{School of Physics and Astronomy, University of Southampton, Southampton SO17 1BJ, United Kingdom}
\author{St\'ephane K\'ena-Cohen}
\email[]{s.kena-cohen@polymtl.ca}
\affiliation{Department of Engineering Physics, \'Ecole Polytechnique de Montr\'eal, Montr\'eal H3C 3A7, QC, Canada}
%\homepage[]{Your web page}
%\thanks{}
%\altaffiliation{}

\date{\today}

\begin{abstract}
We study the use of exciton-polaritons in semiconductor microcavities to generate radiation spanning the infrared to terahertz regions of the spectrum by exploiting transitions between upper and lower polariton branches. The process, which is analogous to difference-frequency generation (DFG), relies on the use of semiconductors with a nonvanishing second-order susceptibility. For an organic microcavity composed of a nonlinear optical polymer, we predict a DFG irradiance enhancement of $2.8\cdot10^2$, as compared to a bare nonlinear polymer film, when triple resonance with the fundamental cavity mode is satisfied. In the case of an inorganic microcavity composed of (111) GaAs, an enhancement of $8.8\cdot10^3$ is found, as compared to a bare GaAs slab. Both structures show high wavelength tunability and relaxed design constraints due to the high modal overlap of polariton modes.
\end{abstract}

% insert suggested keywords - APS authors don't need to do this
%\keywords{}

%\maketitle must follow title, authors, abstract, \pacs, and \keywords
\maketitle

\section{Introduction}

%In addition, the extensive research on exciton polaritons in these systems has led to the experimental observation of a number of nonlinear effects, such as stimulated polariton scattering,\cite{PhysRevLett.84.1547,PhysRevLett.85.3680} parametric oscillation,\cite{PhysRevB.62.R16247,PhysRevLett.85.2793,PhysRevB.63.193305} optical bistability,\cite{PhysRevA.69.023809} condensation and superfluidity.\cite{kasprzak2006bose,amo2009superfluidity}

\begin{figure}[ht]
\includegraphics[width=3.4in]{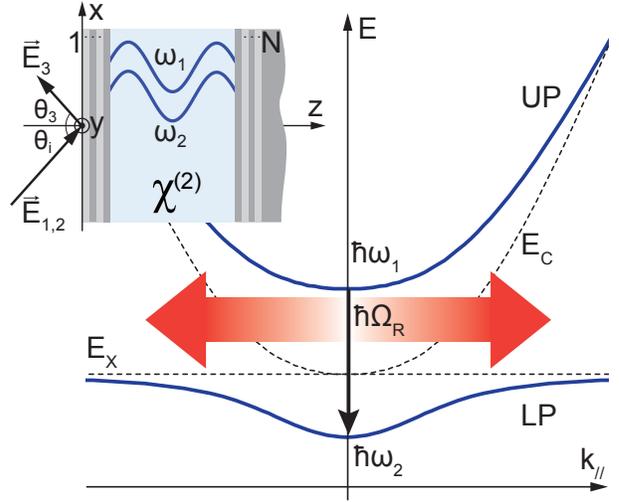}
\caption{\label{figTMM} Dispersion relation of exciton-polaritons as a function of in-plane wavevector. The interaction between an exciton transition ($E_X$) and a Fabry-Perot cavity mode ($E_C$), both represented by dashed lines, leads to the appearance of lower and upper polariton (LP and UP) branches (solid blue). A radiative transition at the Rabi energy ($\hbar\Omega_R$) occurs between two incident pumps at frequencies $\omega_1$ and $\omega_2$ through difference-frequency generation in a second-order nonlinear semiconductor ($\chi^{(2)}\neq0$). Inset: microcavity showing the two pump beams ($\vec{E}_{1,2}$), incident at angle $\theta_i$, and the Rabi radiation ($\vec{E}_3$),  reflected at angle $\theta_3$. The solid blue lines in the $\chi^{(2)}$ layer illustrate the high modal overlap of polariton fields.}
\end{figure}

Half-light, half-matter quasiparticles called polaritons arise in systems where the light-matter interaction strength is so strong that it exceeds the damping due to each bare constituent. In semiconductor microcavities, polaritons have attracted significant attention due to their ability to exhibit strong resonant nonlinearities and to condense into their energetic ground state at relatively low densities. Such polaritons result from the mixing between an exciton transition ($E_X$) and a Fabry-Perot cavity photon ($E_C$). They exhibit a peculiar dispersion, which is shown in Fig.~\ref{figTMM}. Around the degeneracy point of both bare constituents, the lower and upper polariton (LP and UP) branches anticross and their minimum energetic separation is called the vacuum Rabi splitting ($\hbar\Omega_R$). It can range from a few meV in inorganic semiconductors to $\sim1$ eV in organic ones\cite{kasprzak2006bose,PhysRevLett.98.126405,ADOM:ADOM201300256,ph500266d,PhysRevLett.106.196405}. Radiative transitions from the upper to the lower polariton branch can therefore provide a simple route towards tunable infrared (IR) and terahertz (THz) generation.

Such transitions can be understood as resulting from a strongly coupled $\chi^{(2)}$ nonlinear interaction in which two photons, dressed by the resonant interaction with excitons, interact emitting a third photon. As a consequence of the usual $\chi^{(2)}$ selection rule, such polariton-polariton transitions are forbidden in centrosymmetric systems. To overcome this issue several solutions have been proposed, including the use of asymmetric quantum wells,\cite{PhysRevB.87.241304,PhysRevB.89.235309} the mixing of polariton and exciton states with different parity,\cite{PhysRevLett.107.027401,1.3519978} and the use of transitions other than UP to LP.\cite{PhysRevLett.108.197401,PhysRevLett.110.047402}

Here, we study the use of non-centrosymmetric semiconductors, possessing an intrinsic second-order susceptibility $\chi^{(2)}$, to allow for the generation of Rabi-frequency radiation. The irradiance of the resulting UP to LP transitions, which are analogous to classical difference-frequency generation (DFG), have been calculated using a semiclassical model, yielding DFG irradiance enhancements up to almost four orders of magnitude compared to the ones due to the bare $\chi^{(2)}$ nonlinearity. These enhancements can also be related to those expected for parametric fluorescence. Finally, we highlight the use of a triply-resonant scheme to obtain polariton optical parametric oscillation (OPO).

Semiconductor microcavities are advantageous for nonlinear optical mixing due to their ability to spatially and temporally confine the interacting fields. For small interaction lengths, the efficiency of the nonlinear process does not depend on phase-matching, but instead on maximizing the field overlap.\cite{Rodriguez:07} To overcome mode orthogonality, while simultaneously satisfying the symmetry requirements of the $\chi^{(2)}$ tensor, a number of strategies have been proposed such as mode coupling between crossed beam photonic crystal cavities with independently tunable resonances\cite{Burgess:09,1.3607281,Rivoire:11} and the use of single cavities supporting both TE and TM modes.\cite{Zhang:09,1.3568897} Exciton-polaritons provide a simple solution to this problem because they arise from coupling to a single cavity mode and thus naturally display good modal overlap. Many of the fascinating effects observed in strongly-coupled semiconductor microcavities exploit this property, but these have been principally limited to the resonant $\chi^{(3)}$ nonlinearity inherited from the exciton.

Note that in this paper we define the vacuum Rabi frequency as being equal to the resonant splitting due to light-matter coupling. Although this definition is commonly used in the study of quantum light-matter interactions,\cite{Loudon-TheQuantumTheoryOfLight-92} it differs from that often employed in the field of microcavity polaritons,\cite{RevModPhys.85.299} where the vacuum Rabi frequency is defined as being equal to half of the resonant splitting.

This paper is organized as follows. Section \ref{Theory} reviews the nonlinear transfer matrix scheme used to calculate frequency mixing in the small-signal regime. In Sec. \ref{Results}, we calculate the enhancement in irradiance at the Rabi frequency over a bare nonlinear slab for organic and inorganic microcavities and in Sec. \ref{Discussion} we discuss the results and highlight some of the peculiarities of both material sets. Conclusions are presented in Sec. \ref{Conclusion}.

\section{\label{Theory}Theory}

To calculate the propagation of the incident pump fields and the difference-frequency contribution due to nonlinear layers, we use the nonlinear transfer matrix method introduced by \citeauthor{Bethune:89}.\cite{Bethune:89} This method is applicable to structures with an arbitrary number of parallel nonlinear layers,\cite{Liu:13} but is restricted to the undepleted pump approximation, where the three fields are essentially independent. First, we propagate the incident pump fields using the standard transfer matrix method. Within each nonlinear layer, these behave as source terms in the inhomogeneous wave equation. Then, we solve for the particular solution and determine the corresponding source field vectors. Finally, we use the boundary conditions and propagate the free fields using the transfer matrix method to obtain the total field in each layer.

\subsection{\label{TMM}Propagation of the pump fields}

We begin by calculating the field distribution of the two incident pumps as shown in Fig. \ref{figTMM} by using the standard transfer matrix method.\cite{yeh1988optical,born1999principles,1.370757} To simplify the discussion, we consider the pumps to be TE (\^y) polarized. In our notation, the electric field in each layer $i$ is given by sum of two counter-propagating plane waves
\begin{equation}
\mathcal{E}_i^\pm(z,x,t)=\operatorname{Re}\left\{E_i^\pm\thinspace\exp[i(\pm k_{iz}z+k_xx-\omega t)]\right\},
\label{eqE}
\end{equation}
where the $k_{iz}$ and $k_x$ components of the $\vec{k_i}$ wavevector satisfy the relationship $k_{iz}^2+k_x^2=n_i^2(\omega)\,\omega^2/c^2$, with $n_i$ the refractive index of layer $i$. The forward and backward complex amplitudes of the electric field are represented in vector form as $\mathbf{E}_i=\begin{bmatrix}E_i^+&E_i^-\end{bmatrix}^T$. 

%From here on, we omit the $\exp(ik_xx)$ dependence, which remains the same across all interfaces due to the continuity of the tangential (\^x) electric field.

For a given incident field $\mathbf{E}_1$, the field in layer $i$ is calculated by $\mathbf{E}_i=T_i\mathbf{E}_1$, where $T_i$ is the partial transfer matrix
\begin{equation}
T_i=M_{i(i-1)}\phi_{i-1}\cdots M_{21}.
\end{equation}

The interface matrix $M_{ij}$, that relates fields in adjacent interfaces $i$ and $j$, and the propagation matrix $\phi_i$, that relates fields on opposite sides of layer $i$ with thickness $d_i$, are given by
\begin{equation}
M_{ij}=\frac{1}{2k_{iz}}
\begin{bmatrix} 
k_{iz}+k_{jz}&k_{iz}-k_{jz}\\k_{iz}-k_{jz}&k_{iz}+k_{jz}
\end{bmatrix}
\label{eqtm}
\end{equation}
and
\begin{equation}
\phi_i=
\begin{bmatrix} 
\exp(ik_{iz}d_i)&0\\0&\exp(-ik_{iz}d_i)
\end{bmatrix}.
\label{eqphi}
\end{equation}

\subsection{Inclusion of nonlinear polarizations}

To obtain the difference-frequency contribution within a nonlinear layer, we must solve the inhomogeneous wave equation for the electric field
\begin{equation}
\nabla^2\mathcal{E}-\mu\epsilon\frac{\partial^2\mathcal{E}}{\partial t^2}=\mu\frac{\partial^2\mathcal{P}^{NL}}{\partial t^2},
\label{eqWave}
\end{equation}
where the source term
\begin{equation}
\mathcal{P}^{NL}(z,x,t)=\epsilon_0\chi^{(2)}\mathcal{E}^2(z,x,t)
\label{eqPol}
\end{equation}
is the second-order nonlinear polarization, $\mu$ is the magnetic permeability and $\epsilon$ the permittivity. By using a polarization term of the same form as Eq. (\ref{eqE}), Eq.~(\ref{eqWave}) can be written in the frequency domain as
\begin{equation}
\left[-(k^{NL})^2+\omega_{NL}^2n^2(\omega_{NL})\mu_0\epsilon_0\right]\mathbf{E}=-\omega_{NL}^2\mu_0\mathbf{P}^{NL},
\end{equation}
with wavevector $k^{NL}$, $\mu(\omega_{NL})=\mu_0$ and $\epsilon(\omega_{NL})=n^2(\omega_{NL})\epsilon_0$. The nonlinear polarization thus generates a bound source field at the same frequency given by
\begin{equation}
\mathbf{E}_s=\frac{\mathbf{P}^{NL}}{\frac{(k^{NL})^2}{\omega_{NL}^2\mu_0}-n^2(\omega_{NL})\epsilon_0}.
\label{eqES}
\end{equation}

If we consider the presence of two pump fields $\mathbf{E}_1(\omega_1)$ and $\mathbf{E}_2(\omega_2)$, with $\omega_1>\omega_2$, the $\mathcal{E}^2(z,x,t)$ term in Eq. (\ref{eqPol}) can be written as
\begin{equation}
\begin{split}
\mathcal{E}^2(z,x,t)&=\operatorname{Re}\left\{E_1^+\thinspace\exp\left[i\left(k_z^1z+k_x^1x-\omega_1t\right)\right]\right.\\
&\quad\left.+E_1^-\thinspace\exp\left[i\left(-k_z^1z+k_x^1x-\omega_1t\right)\right]\right.\\
&\qquad\left.+E_2^+\thinspace\exp\left[i\left(k_z^2z+k_x^2x-\omega_2t\right)\right]\right.\\
&\qquad\quad\left.+E_2^-\thinspace\exp\left[i\left(-k_z^2z+k_x^2x-\omega_2t\right)\right]\right\}^2.
\end{split}
\end{equation}

Expanding $\mathcal{E}^2(z,x,t)$ leads to terms related to frequency doubling ($\omega_{NL}=2\omega_1$ or $2\omega_2$) and rectification ($\omega_{NL}=0$), sum-frequency ($\omega_{NL}=\omega_1+\omega_2$) and difference-frequency generation ($\omega_{NL}=\omega_1-\omega_2$). The terms contributing to the latter ($\equiv \omega_3$) are given by
\begin{equation}
\begin{split}
\mathcal{P}^3(z,x,t)&=\epsilon_0\chi^{(2)}\operatorname{Re}\left\{\left(E_1^+E_2^{+*}\thinspace\exp\left[i\left(k_z^1-k_z^2\right)z\right]\right.\right.\\
&\quad\left.\left.+E_1^+E_2^{-*}\thinspace\exp\left[i\left(k_z^1+k_z^2\right)z\right]\right.\right.\\
&\qquad\left.\left.+E_1^-E_2^{+*}\thinspace\exp\left[-i\left(k_z^1+k_z^2\right)z\right]\right.\right.\\
&\qquad\quad\left.\left.+E_1^-E_2^{-*}\thinspace\exp\left[-i\left(k_z^1-k_z^2\right)z\right]\right)\right.\\
&\qquad\qquad\left.\times\thinspace\exp\left(i\left[\left(k_x^1-k_x^2\right)x-\omega_3t\right]\right)\right\}.
\end{split}
\end{equation}

Co-propagating waves ($\pm,\pm$) generate terms with perpendicular wavevector $k^{3-}_{z}=k_z^1-k_z^2$, whereas counter-propagating waves ($\pm,\mp$) generate terms with $k^{3+}_{z}=k_z^1+k_z^2$. Their contributions can be handled separately when pump depletion is ignored, so we divide the polarization term into two components 
\begin{subequations}
\begin{align}
\mathbf{P}^{3-}&=\epsilon_0\chi^{(2)}
\begin{bmatrix} 
E_1^+E_2^{+*}\\E_1^-E_2^{-*}
\end{bmatrix}\\  
\mathbf{P}^{3+}&=\epsilon_0\chi^{(2)}
\begin{bmatrix} 
E_1^+E_2^{-*}\\E_1^-E_2^{+*}
\end{bmatrix},
\end{align}
\end{subequations}

with their source fields given by Eq. (\ref{eqES}) and the perpendicular component of $k^{NL}$ taking the values of $k_z^{3-}$ or $k_z^{3+}$, respectively. 

In addition to the bound fields, there are also free fields with frequency $\omega_3$ that are solutions to the homogeneous wave equation. The free field in a nonlinear layer $j$ is obtained from the bound field amplitudes $\mathbf{E}_{js}$ and the boundary conditions at the interfaces. By imposing continuity of the total tangential electric and magnetic fields across interfaces i--j and j--k, an effective free field source vector can be defined as
\begin{equation}
\mathbf{S}_j=\left(\phi_j^{-1}M_{js}\phi_{js}-M_{js}\right)\mathbf{E}_{js},
\label{eqS}
\end{equation}
where the source matrices with the subscript $s$, $M_{js}$ and $\phi_{js}$, are identical to the ones given by Eqs. (\ref{eqtm}) and (\ref{eqphi}), with $k_{iz}$ and $k_{jz}$ taking the values of $k^3_{jz}$ and $k^{3\pm}_{jz}$, respectively.

The total nonlinear field is then given by the sum of independent source field vectors $\mathbf{S}_j$ propagated using the transfer matrix method reviewed in Sec. \ref{TMM}. In particular, for the case where only layer $j$ is nonlinear, we obtain
\begin{equation}
\begin{split}
\begin{bmatrix} 
E_{3T}\\0
\end{bmatrix}
&=M_{N(N-1)}\cdots M_{21}
\begin{bmatrix} 
0\\E_{3R}
\end{bmatrix}\\
&\quad+M_{N(N-1)}\cdots M_{(j+1)j}\mathbf{S}_j\\
&=T_N
\begin{bmatrix} 
0\\E_{3R}
\end{bmatrix}
+
\begin{bmatrix} 
R_j^+\\R_j^-
\end{bmatrix},
\label{eqEk1}
\end{split}
\end{equation}

with
\begin{equation}
\mathbf{R}_j=T_N{T_j}^{-1}\mathbf{S}_j.
\end{equation}

Therefore, the reflected and transmitted components of the $\mathbf{E}_3$ field can be calculated by
\begin{subequations}
\begin{align}
  \label{eqE3R}
  E_{3R}&=-\frac{R_j^-}{T_{22}}\\
  \label{eqE3T}
  E_{3T}&=R_j^+-\frac{T_{12}}{T_{22}}R_j^-.
\end{align}
\end{subequations}

The angle dependence of the reflected difference-frequency field can be expressed as
\begin{equation}
|k_3|\thinspace\sin\theta_3^{\pm}=|k_1|\thinspace\sin\theta_1\pm|k_2|\thinspace\sin\theta_2,
\end{equation}

where the $\pm$ sign must match the wavevector component $k^{3\pm}_z$ when both pumps are incident on the same side of the normal.\cite{PhysRev.128.606} Because the first layer is taken to be air with $n(\omega)=1$, if we consider both pumps to be incident with the same angle $\theta_1=\theta_2=\theta_i$, we obtain for the cases of $k^{3-}_z$ and $k^{3+}_z$
\begin{subequations}
\begin{align}
  \label{eqAng3-}
  \sin\theta_3^{-}&=\frac{\omega_1\thinspace\sin\theta_i-\omega_2\thinspace\sin\theta_i}{\omega_1-\omega_2}=\sin\theta_i\\
  \label{eqAng3+}
  \sin\theta_3^{+}&=\left(\frac{\omega_1+\omega_2}{\omega_1-\omega_2}\right)\thinspace\sin\theta_i.
\end{align}
\end{subequations}

Equation (\ref{eqAng3-}) shows that the DFG component due to co-propagating waves exits the structure at the same angle as the incident pumps, resembling the law of reflection. Conversely, according to Eq. (\ref{eqAng3+}), the component due to counter-propagating pump waves is very sensitive to any angle mismatch between the pumps and easily becomes evanescent for small DFG frequencies.

\section{\label{Results}Results}

\subsection{Organic polymer cavity}

In this section, we investigate the use of organic microcavities for Rabi frequency generation. Due to the large binding energy of Frenkel excitons, organic microcavities can readily reach the strong coupling regime at room temperature and have shown Rabi splittings of up to 1 eV.\cite{ADOM:ADOM201300256,ph500266d} Demonstrations of optical nonlinearities have been more limited than in their inorganic counterparts, but a variety of resonant\cite{PhysRevB.69.235330,PhysRevB.74.113312} and non-resonant nonlinearities\cite{kena2010room,daskalakis2014nonlinear,plumhof2014room} have nevertheless been observed in these systems. 

Although most organic materials possess a negligible second-order susceptibility, a number of poled nonlinear optical (NLO) chromophores have been shown to exhibit high electro-optic coefficients that exceed those of conventional nonlinear crystals such as LiNbO$_3$ by over an order of magnitude.\cite{Boyd,doi:10.1021/cr9000429,doi:10.1021/cr9000429} In addition, the metallic electrodes needed for polling can also be used as mirrors, providing high mode confinement and a means for electrical injection.

We will consider a thin NLO polymer film enclosed by a pair of metallic (Ag) mirrors of thicknesses  10 nm (front) and 100 nm (back). The model polymer is taken to possess a dielectric constant described by a single Lorentz oscillator
\begin{equation}
\epsilon(\omega)=\epsilon_B+\frac{f{\omega_0}^2}{{\omega_0}^2-\omega^2-i\Gamma\omega},
\label{eqDL}
\end{equation}

where $\epsilon_B$ is the background dielectric constant, $f$ is the oscillator strength, $\omega_0$ is the frequency of the optical transition and $\Gamma$ its full width at half maximum (FWHM). The parameters are chosen to be $\epsilon_B=4.62$, $f=0.91$, $\hbar\omega_0=1.55$ eV and $\hbar\Gamma=0.12$ eV. Experimental values are used for the refractive index of Ag.\cite{Rakic:98} For simplicity, we ignore the dispersive nature of the second-order nonlinear susceptibility and take $\chi^{(2)}=300$ pm/V. In principle, the Lorentz model could readily be extended to account for the dispersive resonant behavior.\cite{Boyd}

\begin{figure}[ht]
\includegraphics[width=3.4in]{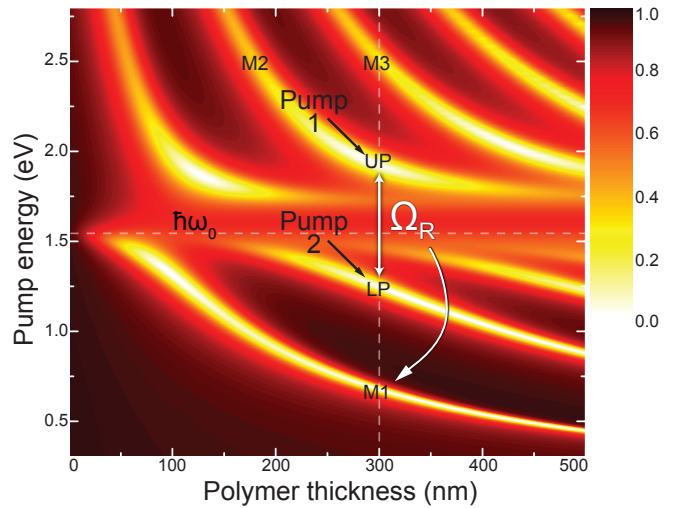}
\caption{\label{figRmap} Reflectance as a function of pump energy and thickness of the polymer film. Front and back Ag mirrors have thicknesses of 10 nm and 100 nm, respectively. Dielectric parameters: $\epsilon_B=4.62$, $f=0.91$, $\hbar\omega_0=1.55$ eV, $\hbar\Gamma=0.12$ eV and $\chi^{(2)}=300$ pm/V. Dashed horizontal line indicates the exciton energy. At the thickness of 300 nm, indicated by a vertical dashed line, the M1 cavity mode is resonant with the difference-frequency generation of pumps 1 and 2 such that $E_{UP}-E_{LP}=\hbar\Omega_R=E_{M1}$.}
\end{figure}

Figure \ref{figRmap} shows the linear reflectance, calculated at normal incidence, as a function of polymer film thickness. The reflectance for film thicknesses below 200 nm shows only the fundamental cavity mode (M1), which is split into UP and LP branches. For these branches, the Rabi energy falls below the LP branch, where there are no further modes available for difference-frequency generation.

By increasing the thickness of the film, low-order modes shift to lower energies and provide a pathway for the DFG radiation to escape. For example, at 300 nm, a triple-resonance condition occurs where the Rabi splitting of the M2 cavity mode matches the M1 energy ($E_{UP}-E_{LP}=\hbar\Omega_R=E_{M1}=0.68$ eV). A second resonance occurs between M3 and LP because $E_{M3}-E_{LP}=E_{LP}=1.25$ eV, but with reduced modal overlap.

\begin{figure}[ht]
\includegraphics[width=3.4in]{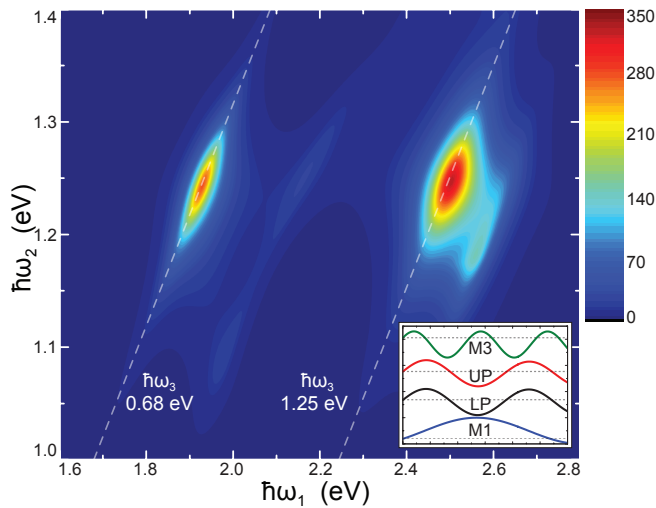}
\caption{\label{figAgDFGmapF} DFG irradiance enhancement of the poled NLO polymer model structure with respect to a bare film of equal thickness. Due to the thickness of the second mirror, only reflected fields are considered. The tilted dashed lines correspond to pairs of pump energies that generate the same DFG energy and that match the M1 (left, $\hbar\omega_3=\hbar\omega_{M1}=0.68$ eV) and LP (right, $\hbar\omega_3=\hbar\omega_{LP}=1.25$ eV) energies in the triple-resonance condition. Inset: normalized electric field profiles of the relevant modes, illustrating the excellent modal overlap of the LP and UP branches.}
\end{figure}

The enhancement in DFG irradiance from the microcavity, as compared to a bare nonlinear slab, is shown in Fig. \ref{figAgDFGmapF} as a function of the pump energies. The two peaks correspond to the triple-resonance conditions mentioned above, where the left peak corresponds to an enhancement of $2.8\cdot10^2$ at the Rabi energy ($\lambda_3=1.82$ $\mu$m) and the right peak to an enhancement of $3.3\cdot10^2$ at the LP energy ($\lambda_{LP}=996$ nm). 

The inset shows the normalized electric field profiles of the relevant modes, which highlight the good modal overlap of the two pump fields in the strong-coupling regime. The small thickness of the front metallic mirror lowers the mutual orthogonality of different modes and accounts for the lack of symmetry of the fields with respect to the center of the film. This loss of orthogonality allows the overlap integral between M3 and LP to be non-zero and the enhanced DFG extraction due to the triple-resonance condition leads to the appearance of the second peak at $\hbar\omega_3=1.25$ eV in Fig. \ref{figAgDFGmapF}. 

\begin{figure}[ht]
\includegraphics[width=3.4in]{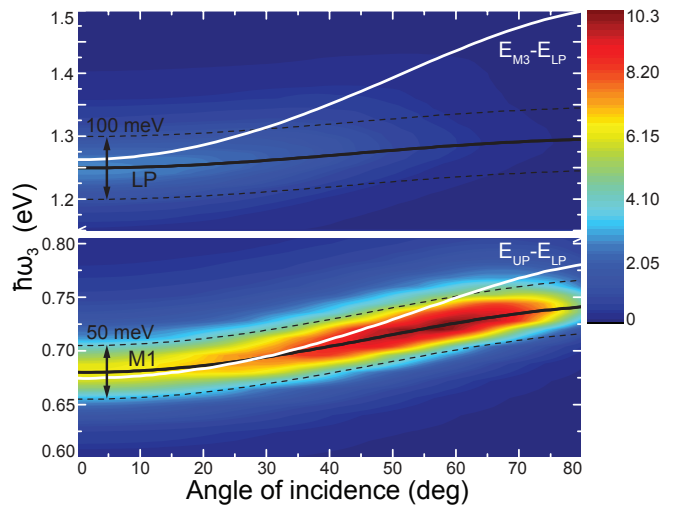}
\caption{\label{figAngles1} Angle dependence of DFG energy and irradiance (kW/m$^2$) for TE polarized pumps incident on the structure with NLO polymer and Ag mirrors when $\theta_1=\theta_2=\theta_i$. Only waves with $k^{3-}_{z}=k_z^1-k_z^2$ are considered. Lower and upper panels show the DFG at the the Rabi and LP energies, respectively. Solid black lines illustrate the energies of the M1 (bottom) and LP (top) modes where DFG radiation can be extracted in triple-resonance. Dashed black lines illustrate a typical linewidth of 100 meV for the LP branch and 50 meV for the M1 mode. Solid white lines indicate the angle dependence of the DFG energy. For the upper panel, as the white line moves out of resonance with the black LP line, the DFG peak is suppressed. For the lower one, a slight increase is observed around 57$^o$ and corresponds to an enhancement of the triple-resonance condition, after which the irradiance rolls off.}
\end{figure}

Additionally, oblique incidence of the pump beams can be used to tune the DFG energy. As indicated by Eq. (\ref{eqAng3+}), the $k^{3+}_z$ component of the DFG signal rapidly becomes evanescent and therefore we shall consider only the $k^{3-}_z$ component. Figure \ref{figAngles1} shows the dependence of DFG energy and irradiance on the angle of incidence when $\theta_1=\theta_2=\theta_i$. In the lower panel, as the interacting modes move to higher energies, the triple-resonance condition at the Rabi ($E_{UP}-E_{LP}$) energy is maintained for incidence angles up to 79$^o$. The maximum irradiance is obtained at 57$^o$ for $\hbar\omega_{NL}=0.72$ eV ($\lambda_{NL}=1.72$ $\mu$m). This enhancement is reduced by 3 dB at $\hbar\omega_3=0.74$ eV ($\lambda_3=1.68$ $\mu$m) for 79$^o$. The upper panel shows that the peak at $\hbar\omega_3=1.25$ eV falls out of the triple resonance condition faster with a 3 dB roll-off at 40$^o$.

\subsection{(111) GaAs cavity}

The vast majority of resonant nonlinearities observed in inorganic semiconductor microcavities are due to a $\chi^{(3)}$ nonlinearity inherited from the exciton.\cite{PhysRevB.62.R4825} In the typical $\chi^{(3)}$ four-wave mixing process, two pump (p) polaritons interact to produce signal (s) and idler (i) components such that their wave-vectors satisfy $2\mathbf{k}_p=\mathbf{k}_s+\mathbf{k}_i$. Second-order susceptibilities tend to be much larger than their $\chi^{(3)}$ counterparts, but conventionally used (001)-microcavities only allow for nonlinear optical mixing between three orthogonally polarized field components.

A number of commonly used inorganic semiconductors are known to be non-centrosymmetric and to possess high second-order susceptibility tensor elements. Examples include III-V semiconductors, such as gallium arsenide (GaAs) and gallium phosphide (GaP), and II-VI semiconductors, such as  cadmium sulfide (CdS) and cadmium selenide (CdSe).\cite{Boyd,shen1984principles} To allow for the nonlinear optical mixing of co-polarized waves to occur, we will consider (111) GaAs as the microcavity material,\cite{1.4833545,Buckley:14} in contrast to the typical (001)-oriented material.

We consider a $\lambda/2$ (111) bulk GaAs microcavity sandwiched between 20 (25) pairs of AlAs/Al$_{0.2}$Ga$_{0.8}$As distributed Bragg reflectors (DBRs) on top (bottom). The structure is followed by a bulk GaAs substrate with the same dielectric constant as the cavity material, modeled by Eq. (\ref{eqDL}) with experimental values $\epsilon_B=12.53$, $f=1.325\cdot10^{-3}$, $\hbar\omega_0=1.515$ eV and $\hbar\Gamma=0.1$ meV.\cite{PhysRevB.52.1800} Experimental values are also used for the refractive index of Al$_x$Ga$_{1-x}$As.\cite{1.336070} The nonlinear susceptibility was kept the same as for the NLO polymer ($\chi^{(2)}=300$ pm/V) to allow for a direct comparison of the irradiances. The absolute value chosen has no effect on the enhancement factor. In practice, the largest contribution to the background $\chi^{(2)}$ in GaAs is due to interband transitions and for simplicity we ignore the resonant contribution to $\chi^{(2)}$.

\begin{figure}[ht]
\includegraphics[width=3.4in]{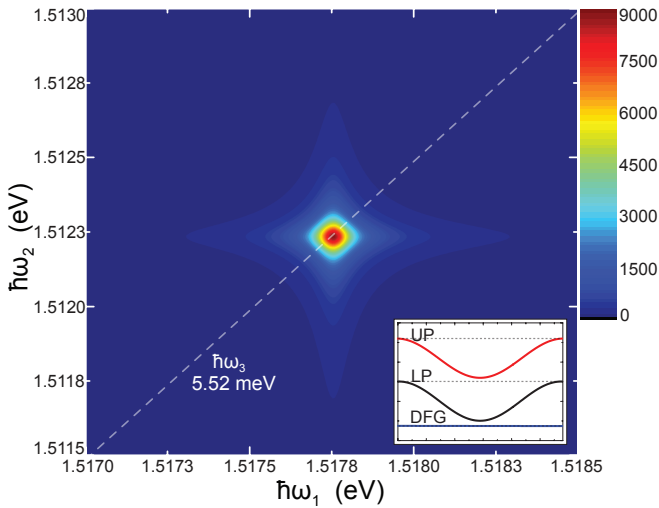}
\caption{\label{figDBRDFGmapF} DFG enhancement of a $\lambda/2$ (111) GaAs cavity structure with respect to a bare slab. GaAs parameters: $\epsilon_B=12.53$, $f=1.325\cdot10^{-3}$, $\hbar\omega_0=1.515$ eV and $\hbar\Gamma=0.1$ meV.\cite{PhysRevB.52.1800} The same value of $\chi^{(2)}=300$ pm/V was used as for the NLO polymer. Due to the presence of the substrate, only reflected fields are considered. The tilted dashed line corresponds to pairs of pump energies that generate the same DFG energy. Inset: normalized electric field profiles inside the GaAs layer illustrating the excellent modal overlap of the LP and UP branches.}
\end{figure}

The enhancement in DFG irradiance as compared to a bare GaAs slab of equal thickness is shown in Fig. \ref{figDBRDFGmapF}. Due to the much smaller oscillator strength in GaAs, as compared to the NLO polymer, the Rabi splitting of $\hbar\omega_3=5.52$ meV falls in the THz range ($\nu_3=1.33$ THz) with an enhancement of $8.8\cdot10^3$.

\begin{figure}[ht]
\includegraphics[width=3.4in]{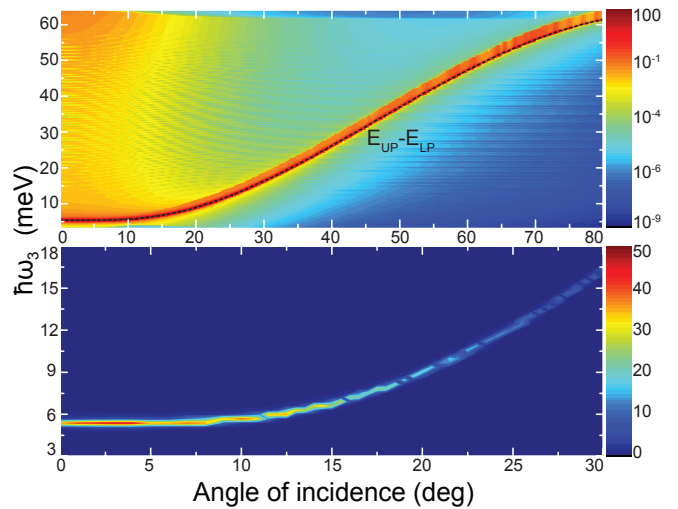}
\caption{\label{figAngles2} Angle dependence of DFG energy and irradiance (W/m$^2$) for TE polarized pumps incident on the $\lambda/2$ (111) GaAs structure with DBR mirrors when $\theta_1=\theta_2=\theta_i$. Only waves with $k^{3-}_{z}=k_z^1-k_z^2$ are considered. The upper panel shows the angle dependence of DFG irradiance in logarithmic scale, with the dashed black line tracing the DFG energy. The lower panel shows a smaller angular range of the same data in linear scale where a fast decrease of DFG irradiance can be observed as the angle of incidence increases.}
\end{figure}

Figure \ref{figAngles2} shows the angle dependence of the DFG energy and irradiance when $\theta_1=\theta_2=\theta_i$. The dashed black line in the upper panel traces the DFG energy, where a logarithmic scale for the irradiance was used due to its rapid decrease with angle of incidence. The lower panel shows a segment of the same data on a linear scale. Tunability down to 3 dB can be obtained up to $\hbar\omega_3=7.21$ meV ($\nu_3=1.74$ THz) at 17$^o$.

\section{\label{Discussion}Discussion}

\begin{figure}[ht]
\includegraphics[width=3.4in]{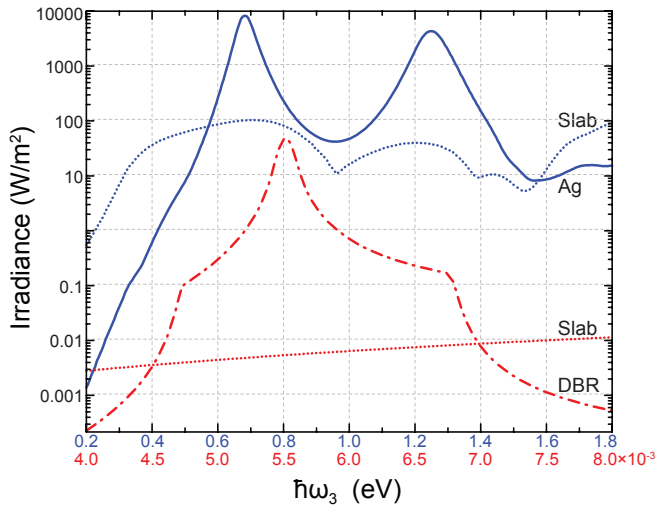}
\caption{\label{figIcomp} Comparison of the calculated DFG irradiances for the two structures studied. Solid blue (dash-dot red) line represents the NLO polymer (GaAs) cavity with Ag (DBR) mirrors and dotted lines directly below represent the corresponding bare slabs. Top blue (bottom red) energy scale relates to the NLO polymer (GaAs) cavity. The curves have been extracted from the maps shown in Fig. \ref{figAgDFGmapF} and Fig. \ref{figDBRDFGmapF} by picking out the maximum values among all pairs of pump energies that generate the same DFG energy. Pump irradiances are $I_1=I_2=10$ GW/m$^2$. }
\end{figure}

In Sec. \ref{Results} we showed that the use of polaritonic modes for Rabi frequency generation can lead to irradiance enhancements of almost four orders or magnitude with respect to bare nonlinear slabs. Quantitative estimates can be obtained by considering equal pump irradiances $I_1=I_2=10$ GW/m$^2$. Figure \ref{figIcomp} shows the maximum DFG irradiances for the two structures and the reference slabs. For the NLO film with Ag mirrors, the calculated peak DFG irradiances are $I_{DFG}=7.69$ kW/m$^2$ at $\hbar\omega_3=0.68$ eV and $I_{DFG}=4.05$ kW/m$^2$ at $\hbar\omega_3=1.25$ eV. For the $\lambda/2$ (111) GaAs microcavity with DBRs, we find $I_{DFG}=45$ W/m$^2$ at $\hbar\omega_3=5.52$ meV.

For the organic microcavity, Fig. \ref{figIcomp} shows that the irradiance due to DFG at the Rabi energy exceeds the one at the LP energy, as expected due to the higher mode overlap. In Fig. \ref{figAgDFGmapF}, however, a higher enhancement was found at the LP energy. This apparent contradiction arises from normalizing each point by the corresponding DFG irradiances of the bare polymer slab. 

There is also a substantial difference in cavity field enhancement for both material sets. Metal losses in the polymer cavity prevent a significant enhancement of the UP and LP electric fields with $|E_{peak}/E_{in}|=1.2$, where $E_{peak}$ and $E_{in}$ are the peak and incident fields, respectively. In contrast, for the GaAs microcavity an enhancement of 15 is obtained. Despite this field enhancement, the irradiance shown in Fig. \ref{figIcomp} is 170 times lower at the Rabi energy for the inorganic microcavity than for the organic one. This is a consequence of the $\omega_{NL}^2$ factor in the source field given by Eq. (\ref{eqES}), making DFG at smaller energies increasingly difficult.

Finally, we can use Fig. \ref{figIcomp} to evaluate the tunability of the structures at normal incidence. For the first structure, the FWHM of the $\hbar\omega_3=0.68$ eV DFG peak is 0.045 eV, indicating that the same structure can be used for DFG generation from 1.76 $\mu$m to 1.88 $\mu$m by adjustment of the pumps only. For the GaAs structure, the FWHM of the $\hbar\omega_3=5.52$ meV DFG peak is 0.12 meV, indicating a tunability from $\nu_3=1.32$ THz to $\nu_3=1.35$ THz. 

We should note that although in our calculation two pumps were used, similar enhancements are anticipated for (spontaneous) parametric fluorescence ($I_2=0$). In addition, the triply-resonant scheme introduced for the organic microcavity where the signal is resonant has further consequences. First, coupled-mode theory analysis of triply-resonant systems has shown the existence of critical input powers to maximize nonlinear conversion efficiency.\cite{Burgess:09,Burgess:09OE} These are found to be inversely proportional to the product of the Q-factors. Lower Q-factors are thus advantageous for high power applications. Second, the scheme is also well-suited for realizing a more conventional $\chi^{(2)}$ polariton OPO.  In this case, the oscillation threshold can be shown to depend inversely on the product of Q-factors.

Since in general, any $\chi^{(2)}$ medium will also have a non-zero $\chi^{(3)}$, these structures will display a change in refractive index proportional to the square of the applied electric field, an effect known as self/cross-phase modulation. The power dependance of the refractive index can lead to rich dynamics such as multistability and limit-cycle solutions.\cite{PhysRevA.83.033834,PhysRevA.89.053839}

\section{\label{Conclusion}Conclusion}

We studied the potential for generating Rabi-frequency radiation in microcavities possessing a non-vanishing second-order susceptibility. Using a semiclassical model based on nonlinear transfer matrices in the undepleted pump regime, we calculated the Rabi splitting and the DFG irradiance enhancement for an organic microcavity, composed of a poled nonlinear optical polymer, and for an inorganic one, composed of GaAs. In the first case, we obtained a Rabi splitting of $\hbar\omega_3=0.68$ eV ($\lambda_3=1.82$ $\mu$m) and an enhancement of two orders of magnitude, as compared to a bare polymer film. In the second case, we found a Rabi splitting of $\hbar\omega_3=5.52$ meV ($\nu_3=1.33$ THz) and an enhancement of almost four orders of magnitude, as compared to a bare GaAs slab. These results show the potential of the use of polaritonic modes for IR and THz generation. Both model structures display a high degree of frequency tunability by changing the wavelength and angle of incidence of the incoming pump beams. Similar enhancements are anticipated for parametric fluorescence and the triply-resonant scheme introduced for the optical microcavity can be exploited to realize monolithic $\chi^{(2)}$ OPOs.\\

\begin{acknowledgments}
FB and SKC acknowledge support from the Natural Sciences and Engineering Research Council of Canada.
SDL acknowledges support from the Engineering and Physical Sciences Research Council (EPSRC), research grant EP/L020335/1. SDL is Royal Society Research Fellow. 	
\end{acknowledgments}

\bibliography{references}

\end{document}